\documentclass[english,aps,preprint]{revtex4}
\usepackage[T1]{fontenc}
\usepackage[latin9]{inputenc}
\usepackage{array}
\usepackage{multirow}
\usepackage{amsmath}
\usepackage{amssymb}
\usepackage{graphicx}

\makeatletter

\providecommand{\tabularnewline}{\\}

\@ifundefined{textcolor}{}
{%
 \definecolor{BLACK}{gray}{0}
 \definecolor{WHITE}{gray}{1}
 \definecolor{RED}{rgb}{1,0,0}
 \definecolor{GREEN}{rgb}{0,1,0}
 \definecolor{BLUE}{rgb}{0,0,1}
 \definecolor{CYAN}{cmyk}{1,0,0,0}
 \definecolor{MAGENTA}{cmyk}{0,1,0,0}
 \definecolor{YELLOW}{cmyk}{0,0,1,0}
}

\makeatother

\usepackage{babel}
\begin{document}

\title{Pair Production of New Heavy Leptons with $U(1)'$ Charge at Linear
Colliders}

\author{V. Ar\i{}}

\email{vari@science.ankara.edu.tr}

\selectlanguage{english}%

\affiliation{Department of Physics, Ankara University, 06100, Ankara, Turkey}

\author{O. Çak\i{}r}

\email{ocakir@science.ankara.edu.tr}

\selectlanguage{english}%

\affiliation{Department of Physics, Ankara University, 06100, Ankara, Turkey}

\author{S. Kuday}

\email{kuday@science.ankara.edu.tr}

\selectlanguage{english}%

\affiliation{Institute of Accelerator Technologies, Ankara University, 06830,
Ankara, Turkey}
\begin{abstract}
We study the pair production of new heavy leptons within a new $U(1)'$
symmetry extension of the Standard Model. Because of the new symmetry,
the production and decay modes of the new heavy leptons would be different
from those of three families of the standard model. The pair production
cross sections depending on the mixing parameter and the mass of heavy
leptons have been calculated for the center of mass energies of $0.5$
TeV, $1$ TeV and $3$ TeV. The accessible ranges of the parameters
have been obtained for different luminosity projections at linear
colliders. We find the sensitivity to the range of mixing parameter
$-1<x<1$ for the mass range $M_{l'}<800$ GeV at $\sqrt{s}=3$ TeV
and $L_{int}=100$ fb$^{-1}$
\end{abstract}
\maketitle

\section{Introduction}

The standard model (SM) of particle physics has been shown to successfully
describe fundamental particle interactions. However, some of the problems
(such as symmetry breaking, dark matter, flavor and CP violation)
do not find adequate answers within the standard model of three fermion
families, and there are considerable interest in new heavy fermions
to be searched at high energy colliders. On the quark sector, the
direct searches performed by the ATLAS collaboration excludes the
heavy quark masses $m_{b'}\lesssim670$ GeV and $m_{t'}\lesssim656$
GeV \cite{ATLAS}, and CMS collaborations have ruled out new heavy
down-type and up-type quarks with masses $m_{b'}\lesssim611$ GeV
and $m_{t'}\lesssim570$ GeV \cite{CMS}. In these experimental bounds,
the branching of decay $b'\to tW$ or $t'\to bW$ is assumed to be
unity; relaxing them leads to slightly weaker bounds as discussed
in \cite{Flacco2010}. On the lepton sector, the LEP measurement of
the $Z$ boson width strongly supports the fact there are only three
light active neutrinos \cite{Nakamura2010}. This measurement leads
to a mass constraint for a new heavy stable neutrino of mass $m_{\nu'}>45$
GeV. In addition, direct production searches at LEPII establish a
lower limit of the order of $100$ GeV for a new charged lepton ($l'$)
and unstable neutrino. It is also straightforward to check that the
recent discovery of Higgs boson with mass of $125$ GeV tune the new
neutrino mass $m_{\nu'}>m_{H}/2$. Furthermore, the off-diagonal lepton
mixings are required to be smaller than 0.115 consistent with the
recent global fits \cite{Lacker2010}. 

One of the simplest extensions of the SM is to include an $U(1)'$
gauge symmetry. After the $U(1)'$ symmetry breaking, there could
remain a residual discrete symmetry as emphasized in Ref. \cite{Soni2013},
and this would cause the lightest new fermion to be stable. The discrete
symmetry for the new model could protect new heavy fermions from the
SM fermions to explain $Z$ boson width measurement at LEP. The model
could provide a stable particle for dark matter candidate, a new source
of CP violation, baryon$-$lepton number conservation. The particle
spectrum of the model \cite{Soni2013} will include, in addition to
the SM particles, an extra heavy fermion family, an extra Higgs doublet,
a Higgs singlet as well as an extra gauge boson ($Z'$). Since the
SM fermions have vanishing $U(1)'$ charges, the new gauge boson $Z'$
cannot form a dilepton or dijet signal besides the $Z-Z'$ mixing
effects which is constrained to be tiny by the precise $Z$ measurements
at LEP \cite{Langacker2009}. Recently, the most stringent limits
on a heavy neutral gauge boson $Z'_{S}$ with the same universal couplings
to fermions as the $Z$ boson (sequential model) have been set using
measurements from ATLAS \cite{ATLAS_2013_ZP} and CMS \cite{CMS_2012_ZP},
translated to a lower bound on the mass $m_{Z'}>2.79$ TeV and $2.96$
TeV, respectively.

In this work, we use the new heavy fermion model accompanied by an
$U(1)'$ symmetry under which only the heavy fermions have nonzero
charge. The heavy charged lepton pair production at Linear Colliders
(LC) through the process $e^{+}e^{-}\to l^{'+}l^{'-}$ and subsequent
decay of $l^{'-}\to\nu'W^{-}$ ending up with a stable heavy neutrino
have been examined at linear collider energies $\sqrt{s}=0.5$ TeV
for the International Linear Collider (ILC) \cite{ILC2013}, $1$
TeV for its upgrade and $3$ TeV for Compact Linear Collider (CLIC)
\cite{CLIC2013}.

\section{Heavy Leptons}

The interactions of new heavy leptons ($l'$,$\nu'$) can be described
by the following Lagrangian 

\begin{eqnarray*}
L' & = & -g_{e}\bar{l'}\gamma^{\mu}l'A_{\mu}-\frac{g_{z}}{2}\bar{l'}\gamma^{\mu}(c_{V}^{l'}-c_{A}^{l'}\gamma^{5})l'Z_{\mu}-\frac{g_{z}}{2}\bar{\nu'}\gamma^{\mu}(c_{V}^{\nu'}-c_{A}^{\nu'}\gamma^{5})\nu'Z_{\mu}\\
 &  & \frac{g_{z}}{2}\bar{l'}\gamma^{\mu}(c'{}_{V}^{l'}-c'{}_{A}^{l'}\gamma^{5})l'Z'_{\mu}-\frac{g_{z}}{2}\bar{\nu'}\gamma^{\mu}(c'{}_{V}^{\nu'}-c'{}_{A}^{\nu'}\gamma^{5})\nu'Z'_{\mu}\\
 &  & -\frac{g_{w}}{2\sqrt{2}}U_{\nu'l'}\bar{l'}\gamma^{\mu}(1-\gamma^{5})\nu'W_{\mu}+H.c.
\end{eqnarray*}
where $g_{e}$, $g_{W}$ and $g_{Z}$ are the electromagnetic, weak-charged
and weak-neutral coupling constants, respectively. The $A_{\mu}$,
$W_{\mu}$ and $Z_{\mu}$ are the fields for photon, $W$ boson and
$Z$ boson, respectively. The field $Z'_{\mu}$ is for the new $Z'$
boson. The $U_{\nu'l'}$ is the mixing element for the charged current
coupling of heavy leptons. Here, we consider a long-lived neutrino
with unit mixing element. The $c_{V}$ and $c_{A}$ are vector and
axial-vector couplings with the $Z$ boson. The $c'_{V}$ and $c'_{A}$
are vector and axial-vector couplings with the new $Z'$ boson. These
couplings can be expressed as $c_{V}={\cal Q}_{L}+{\cal Q}_{R}$ and
$c_{A}={\cal Q}_{L}-{\cal Q}_{R}$ with the left and right handed
fermion gauge charges ${\cal Q}_{L}$ and ${\cal Q}_{R}$, respectively.
The $U(1)'$ charge is defined as ${\cal Q}=(B-L)+xY$ with the mixing
parameter $x$. In the model, the SM fermions are not charged under
the $U(1)'$ while the new fermions have the gauge charges as shown
in Table \ref{tab:tab1}.

\begin{table}
\caption{The $U(1)'$ charges of new heavy fermions with a new parity. \label{tab:tab1}}

\begin{tabular}{|c|c|c|c|}
\hline 
Field & $U(1)'$ charge & $c_{V}$ & $c_{A}$\tabularnewline
\hline 
$t'_{L}$ & $1/3+x$ & \multirow{2}{*}{$2/3+5x$} & \multirow{2}{*}{$-3x$}\tabularnewline
\cline{1-2} 
$t'_{R}$ & $1/3+4x$ &  & \tabularnewline
\hline 
$b'_{L}$ & $1/3+x$ & \multirow{2}{*}{$2/3-x$} & \multirow{2}{*}{$3x$}\tabularnewline
\cline{1-2} 
$b'_{R}$ & $1/3-2x$ &  & \tabularnewline
\hline 
$l'_{L}$ & $-1-3x$ & \multirow{2}{*}{$-2-9x$} & \multirow{2}{*}{$3x$}\tabularnewline
\cline{1-2} 
$l'_{R}$ & $-1-6x$ &  & \tabularnewline
\hline 
$\nu'_{L}$ & $-1-3x$ & \multirow{2}{*}{$-2-3x$} & \multirow{2}{*}{$-3x$}\tabularnewline
\cline{1-2} 
$\nu'_{R}$ & $-1$ &  & \tabularnewline
\hline 
\end{tabular}
\end{table}

As it can be seen from Table \ref{tab:tab1}, the key values for the
parameter $x$ can be calculated as follows: for quarks $c_{V}^{t'}(c_{V}^{b'})$
vanishes when $x=-2/15(2/3)$, respectively; and for leptons $c_{V}^{l'}(c_{V}^{\nu'})$
vanishes when $x=-2/9(-2/3)$, which are in the range of $-2/3<x<2/3$.
A specific value of the mixing parameter $x=0$ corresponds to a vector-type
coupling with the $Z$ boson. Here, we assume that charged heavy lepton
decays through the process $l'\to\nu'+W^{-}$ within a large range
of parameters (mixing and mass). If the mass difference between the
heavy lepton and neutrino is small enough, then the heavy lepton decay
may result in missing transverse energy and off-shell $W$ boson.

\section{Cross Sections}

\begin{figure}
\includegraphics{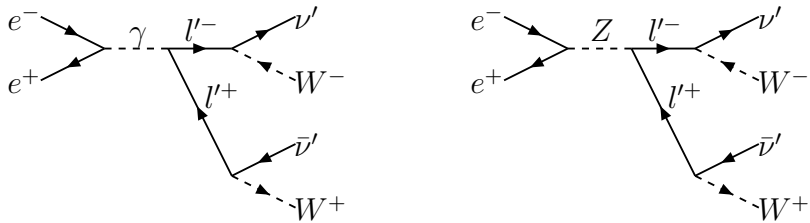}

\caption{Diagrams contributing to the pair production process at linear colliders.
\label{fig:fig1}}
\end{figure}

The pair production of heavy charged leptons can be performed through
the process $e^{+}e^{-}\to l^{'+}l^{'-}$ and subsequent decays of
$l^{'-}\to\nu'W^{-}$ and $l^{'+}\to\bar{\nu}'W^{+}$ ending up with
a pair of stable heavy neutrinos and a pair of $W$ bosons. The contributing
diagrams for the signal are shown in Figure \ref{fig:fig1}. The calculation
of the cross sections for the signal and background is performed using
CalcHEP \cite{Belyaev2013} with the implementation of new heavy leptons
and their interactions. 

\begin{figure}
\includegraphics{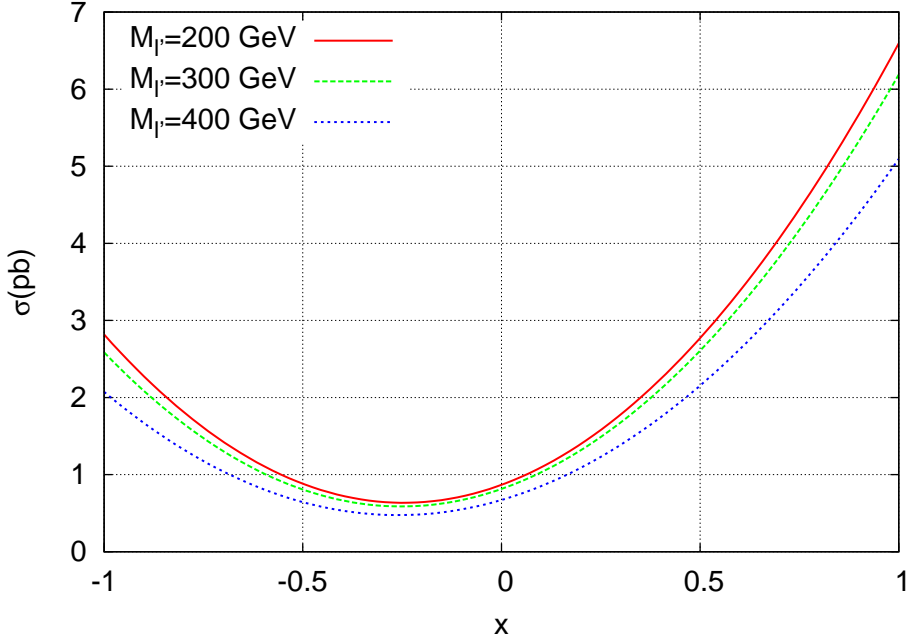}

\caption{The pair production cross sections for the process $e^{+}e^{-}\to l'^{+}l'^{-}$
depending on the parameter $x$ for different mass values of $M_{l'}=200$,
$300$ and $400$ GeV at the center of mass energy $\sqrt{s}=1$ TeV.
\label{fig:fig2}}
\end{figure}

\begin{figure}
\includegraphics{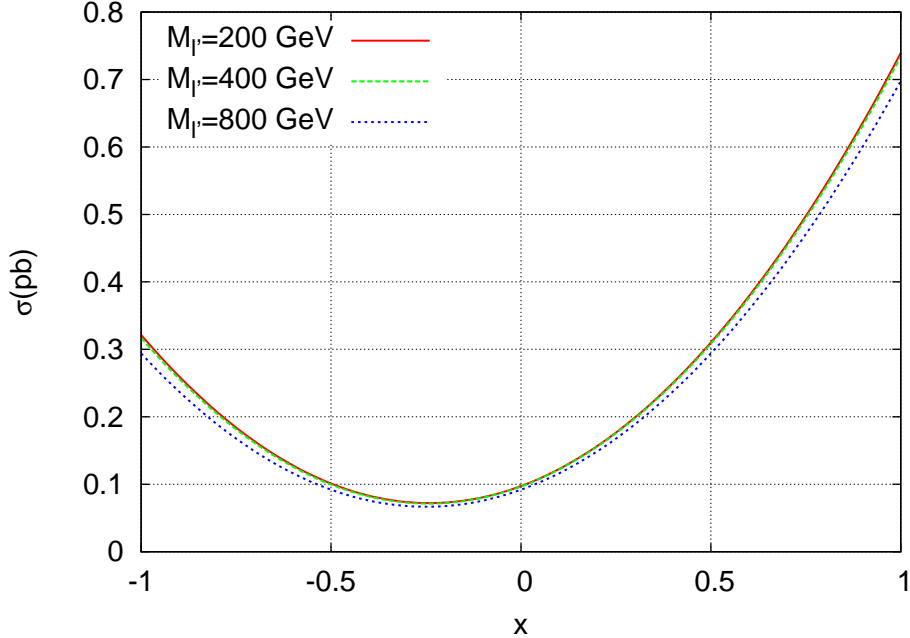}

\caption{The same as Fig. \ref{fig:fig2}, but for the mass values of $M_{l'}=200$,
$400$ and $800$ GeV at $\sqrt{s}=3$ TeV. \label{fig:fig3}}
\end{figure}

At $\sqrt{s}=0.5$ TeV, the cross sections for the signal of heavy
lepton pairs with mass $M_{l'}=200$ GeV are given as $8.72$, $1.38$,
$0.34$, $1.06$, $3.52$, $7.74$ and $21.40$ pb for the parameter
$x$ values $-1.0$, $-0.5$, $-0.25$, $0.0$, $0.25$, $0.50$ and
$1.0$, respectively. The $e^{+}e^{-}$ collider (ILC) with $\sqrt{s}=0.5$
TeV, has the potential up to the kinematical range ($m_{l',\nu'}\leq250$
GeV) for the direct production of new heavy lepton pairs. However,
the CLIC with multi-TeV extends the mass range for the new heavy leptons.
At the center of mass energies of the linear colliders with $\sqrt{s}=1$
TeV and $\sqrt{s}=3$ TeV, the cross sections for the signal are shown
in Figure \ref{fig:fig2} and Figure \ref{fig:fig3} , respectively.
It is clear from Figures \ref{fig:fig2} and \ref{fig:fig3}, the
cross section has a minimum around $x=-0.25$. At the center of mass
energy $\sqrt{s}=1$ (3) TeV, we calculate the change in the cross
sections as $\Delta\sigma\simeq0.5$ ($0.05$) pb for a change in
mixing parameter $|\Delta x|=0.25$ around the minimum. The signal
cross sections depending on the heavy lepton mass and mixing parameter
are given in Table \ref{tab:tab2} for $\sqrt{s}=1$ TeV and in Table
\ref{tab:tab3} for $\sqrt{s}=3$ TeV.

\begin{table}
\caption{Cross sections (pb) for the signal process $e^{+}e^{-}\to l'^{+}l'^{-}$
at $\sqrt{s}=1$ TeV depending on the heavy lepton mass $M_{l'}$
and the parameter $x$. \label{tab:tab2}}

\begin{tabular}{|c|c|c|c|c|c|c|c|}
\hline 
 & \multicolumn{7}{c|}{$x$}\tabularnewline
\hline 
$M_{l'}(GeV)$  & $-1.0$ & $-0.5$ & $-0.25$ & $0.0$ & $0.25$ & $0.5$ & $1.0$\tabularnewline
\hline 
$200$ & $2.82$ & $4.71\times10^{-1}$ & $1.21\times10^{-1}$ & $3.19\times10^{-1}$ & $1.07$ & $2.36$ & $6.59$\tabularnewline
\hline 
$300$ & $2.59$ & $4.24\times10^{-1}$ & $1.09\times10^{-1}$ & $3.04\times10^{-1}$ & $1.01$ & $2.23$ & $6.19$\tabularnewline
\hline 
$400$ & $2.07$ & $3.32\times10^{-1}$ & $8.55\times10^{-2}$ & $2.55\times10^{-1}$ & $0.84$ & $1.84$ & $5.10$\tabularnewline
\hline 
\end{tabular}
\end{table}

\begin{table}
\caption{Cross sections (pb) for the signal process $e^{+}e^{-}\to l'^{+}l'^{-}$
at $\sqrt{s}=3$ TeV depending on the heavy lepton mass $M_{l'}$
and the parameter $x$. \label{tab:tab3}}

\begin{tabular}{|c|c|c|c|c|c|c|c|}
\hline 
 & \multicolumn{7}{c}{$x$}\tabularnewline
\hline 
$M_{l'}(GeV)$  & $-1.0$ & $-0.5$ & $-0.25$ & $0.0$ & $0.25$ & $0.5$ & $1.0$\tabularnewline
\hline 
$200$ & $3.21\times10^{-1}$ & $5.47\times10^{-2}$ & $1.42\times10^{-2}$ & $3.55\times10^{-2}$ & $1.19\times10^{-1}$ & $2.64\times10^{-1}$ & $7.39\times10^{-1}$\tabularnewline
\hline 
$300$ & $3.20\times10^{-1}$ & $5.43\times10^{-2}$ & $1.40\times10^{-2}$ & $3.54\times10^{-2}$ & $1.19\times10^{-1}$ & $2.63\times10^{-1}$ & $7.37\times10^{-1}$\tabularnewline
\hline 
$400$ & $3.17\times10^{-1}$ & $5.37\times10^{-2}$ & $1.39\times10^{-2}$ & $3.54\times10^{-2}$ & $1.18\times10^{-1}$ & $2.62\times10^{-1}$ & $7.34\times10^{-1}$\tabularnewline
\hline 
$500$ & $3.13\times10^{-1}$ & $5.28\times10^{-2}$ & $1.37\times10^{-2}$ & $3.53\times10^{-2}$ & $1.18\times10^{-1}$ & $2.61\times10^{-1}$ & $7.29\times10^{-1}$\tabularnewline
\hline 
$600$ & $3.09\times10^{-1}$ & $5.18\times10^{-2}$ & $1.34\times10^{-2}$ & $3.51\times10^{-2}$ & $1.17\times10^{-1}$ & $2.59\times10^{-1}$ & $7.22\times10^{-1}$\tabularnewline
\hline 
$700$ & $3.02\times10^{-1}$ & $5.04\times10^{-2}$ & $1.31\times10^{-2}$ & $3.48\times10^{-2}$ & $1.16\times10^{-1}$ & $2.55\times10^{-1}$ & $7.12\times10^{-1}$\tabularnewline
\hline 
$800$ & $2.94\times10^{-1}$ & $4.87\times10^{-2}$ & $1.26\times10^{-2}$ & $3.43\times10^{-2}$ & $1.14\times10^{-1}$ & $2.51\times10^{-1}$ & $6.98\times10^{-1}$\tabularnewline
\hline 
\end{tabular}
\end{table}

The cross sections for the background processes contributing to the
lepton+dijet and missing transverse energy (MET) in the the final
state are given in Table \ref{tab:tab4}. The cross sections for the
processes can be multiplied by the corresponding branching ratios,
such as $BR(W^{+}\to l^{+}\nu)=0.22$, $BR(W^{+}\to2j)=0.68$, $BR(Z\to\nu\bar{\nu})=0.20$
and $BR(h\to ZZ^{*})=0.016$ to obtain the cross sections for the
interested final state.

\begin{table}
\caption{Cross sections (pb) for the dominant background contributing to the
final state $l^{\pm}+2j+\mbox{MET}$. \label{tab:tab4}}

\begin{tabular}{|c|c|c|c|}
\hline 
Cross sections (pb) & $\sqrt{s}=0.5$ TeV & $\sqrt{s}=1$ TeV & $\sqrt{s}=3$ TeV\tabularnewline
\hline 
$e^{+}e^{-}\to W^{+}W^{-}Z$ & $4.39\times10^{-2}$ & $6.52\times10^{-2}$ & $3.82\times10^{-2}$\tabularnewline
\hline 
$e^{+}e^{-}\to W^{+}W^{-}h$ & $5.85\times10^{-3}$ & $4.21\times10^{-3}$ & $1.25\times10^{-3}$\tabularnewline
\hline 
$e^{+}e^{-}\to W^{+}W^{-}$ & $7.68\times10^{0}$ & $2.87\times10^{0}$ & $4.98\times10^{-1}$\tabularnewline
\hline 
$e^{+}e^{-}\to W^{+}W^{-}\nu\bar{\nu}$ & $1.09\times10^{-2}$ & $3.38\times10^{-2}$ & $1.47\times10^{-1}$\tabularnewline
\hline 
$l^{\pm}+2j+\mbox{MET}$ & $1.15\times10^{0}$ & $4.36\times10^{-1}$ & $9.76\times10^{-2}$\tabularnewline
\hline 
\end{tabular}
\end{table}

\section{Analysis}

\begin{figure}
\includegraphics{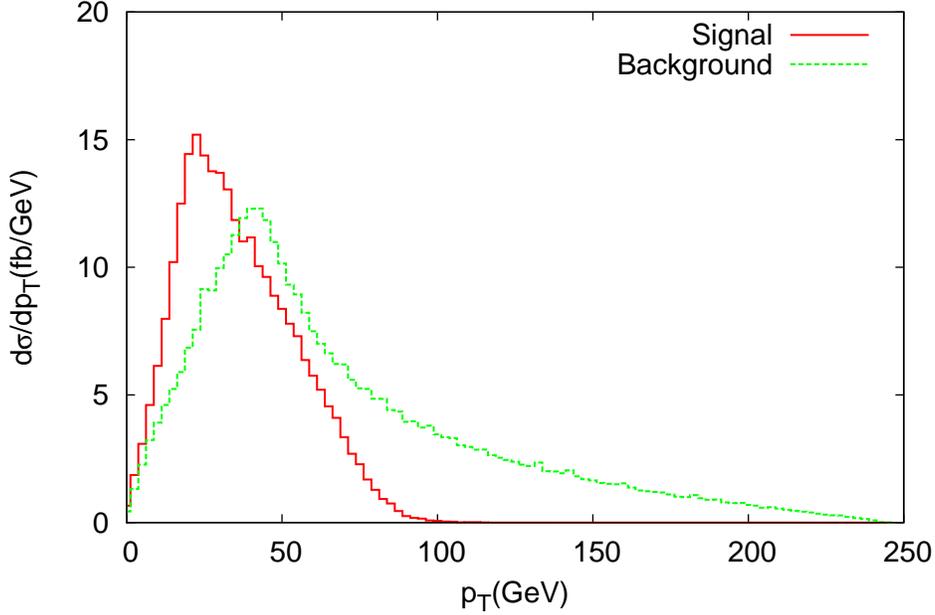}

\caption{The transverse momentum distribution of the final state muon ($\mu^{-}$)
from the signal process $e^{+}e^{-}\to W^{+}\mu^{-}\bar{\nu}_{\mu}\nu'\bar{\nu}'$
with the parameters $x=1$ and $M_{l'}=200$ GeV, and from the background
process $e^{+}e^{-}\to W^{+}\mu^{-}\bar{\nu}_{\mu}$ at $\sqrt{s}=500$
GeV. \label{fig:fig4}}
\end{figure}

\begin{figure}
\includegraphics{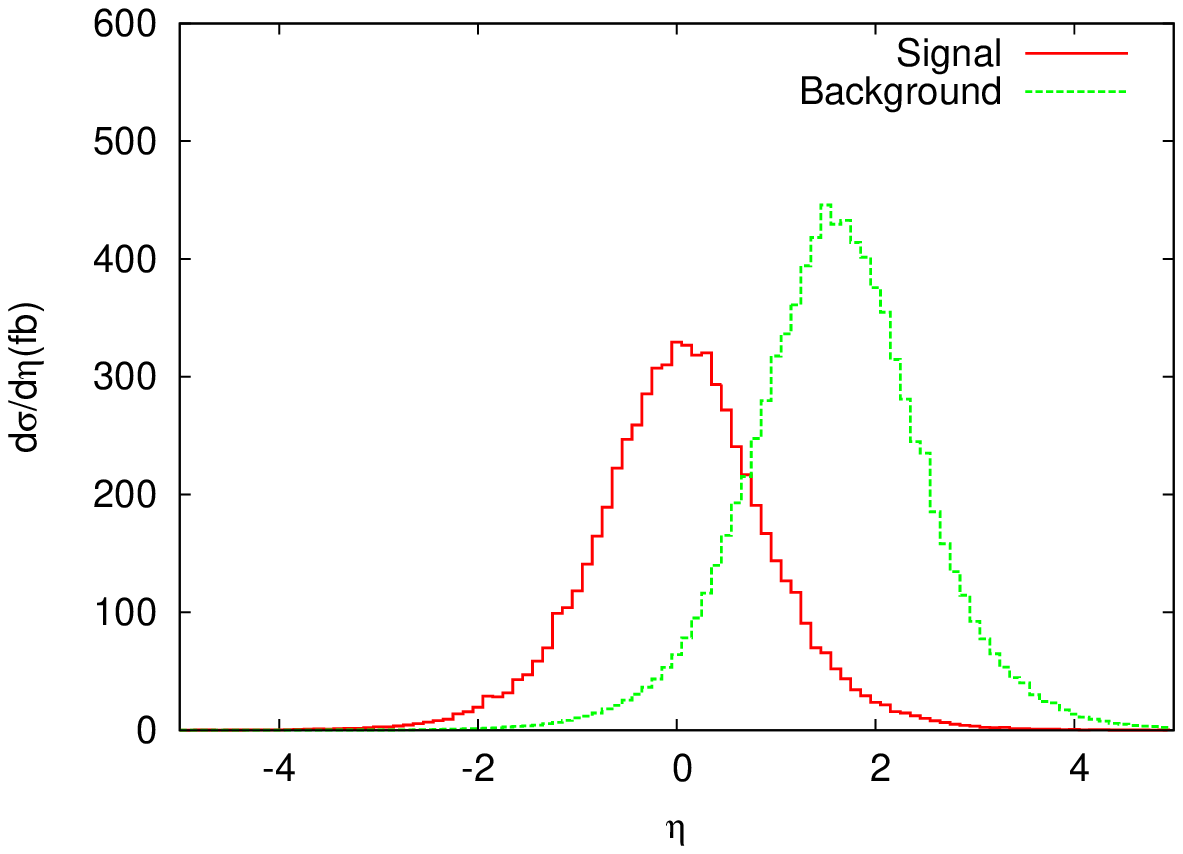}

\caption{The pseudo-rapidty distribution of the final state muon ($\mu^{-}$)
from the signal process $e^{+}e^{-}\to W^{+}\mu^{-}\bar{\nu}_{\mu}\nu'\bar{\nu}'$
with the parameters $x=1$ and $M_{l'}=200$ GeV and from background
process $e^{+}e^{-}\to W^{+}\mu^{-}\bar{\nu}_{\mu}$ at $\sqrt{s}=500$
GeV. \label{fig:fig5}}
\end{figure}

We take into account the final state containing $l^{\pm}+2j+$MET.
The transverse momentum and pseudo-rapidity distributions of the final
state lepton from the signal and background process are given in Fig.
\ref{fig:fig4} and \ref{fig:fig5}, respectively. It is seen from
Fig. \ref{fig:fig4} that a transverse momentum cut $p_{T}>15$ GeV
on the lepton will be required for the acceptance. However, higher
$p_{T}$ cut will not improve the signal to background ratio in the
analysis. Furthermore, the lepton pseudo-rapidity cut of $-1.5<\eta<1.5$
can be used to suppress the relevant background as seen from Fig.
\ref{fig:fig5}. The difference between the pseudo-rapidity distributions
for the signal and background is more pronounced at higher center
of mass energies such as $\sqrt{s}=1$ TeV and $3$ TeV. The cross
section for the main background process $e^{+}e^{-}\to W^{+}W^{-}$
with $W^{\pm}\to l^{\pm}\bar{\nu}_{l}$ and $W^{\mp}\to q\bar{q}'$
will be reduced by $60\%$, $75\%$ and $90\%$, after applying these
cuts, depending on the center of mass energies $\sqrt{s}=0.5$, $1$
and $3$ TeV, respectively. Moreover, the other background process
$e^{+}e^{-}\to W^{+}W^{-}Z$ (where $W^{\pm}\to l^{\pm}\bar{\nu}_{l}$,
$W^{\mp}\to q\bar{q}'$ and $Z\to\nu_{l}\bar{\nu}_{l}$) will give
contribution with a reduction in cross section by $50\%$ after the
cuts.

\begin{figure}
\includegraphics[scale=0.9]{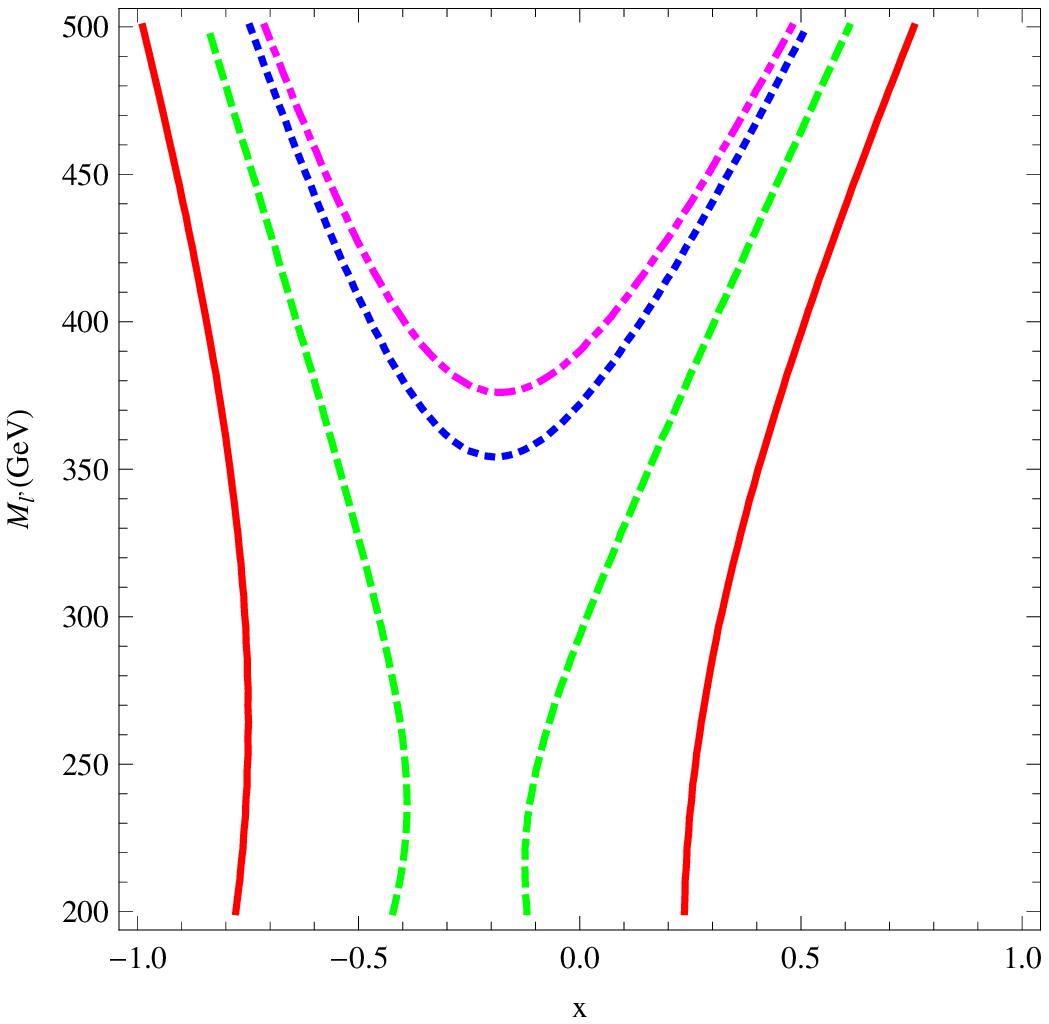}

\caption{Exclusion plot ($95\%$ C.L.) for the heavy lepton mass $M_{l'}$
and mixing parameter $x$ at $\sqrt{s}=1$ TeV and different integrated
luminosities: thick (red) line, dashed (green) line, dotted (blue)
line and dot-dashed (magenta) line corresponds to $L_{int}=20$, $100$,
$1000$ pb$^{-1}$ and $10$ fb$^{-1}$, respectively. \label{fig:fig6}}

\end{figure}

\begin{figure}
\includegraphics[scale=0.9]{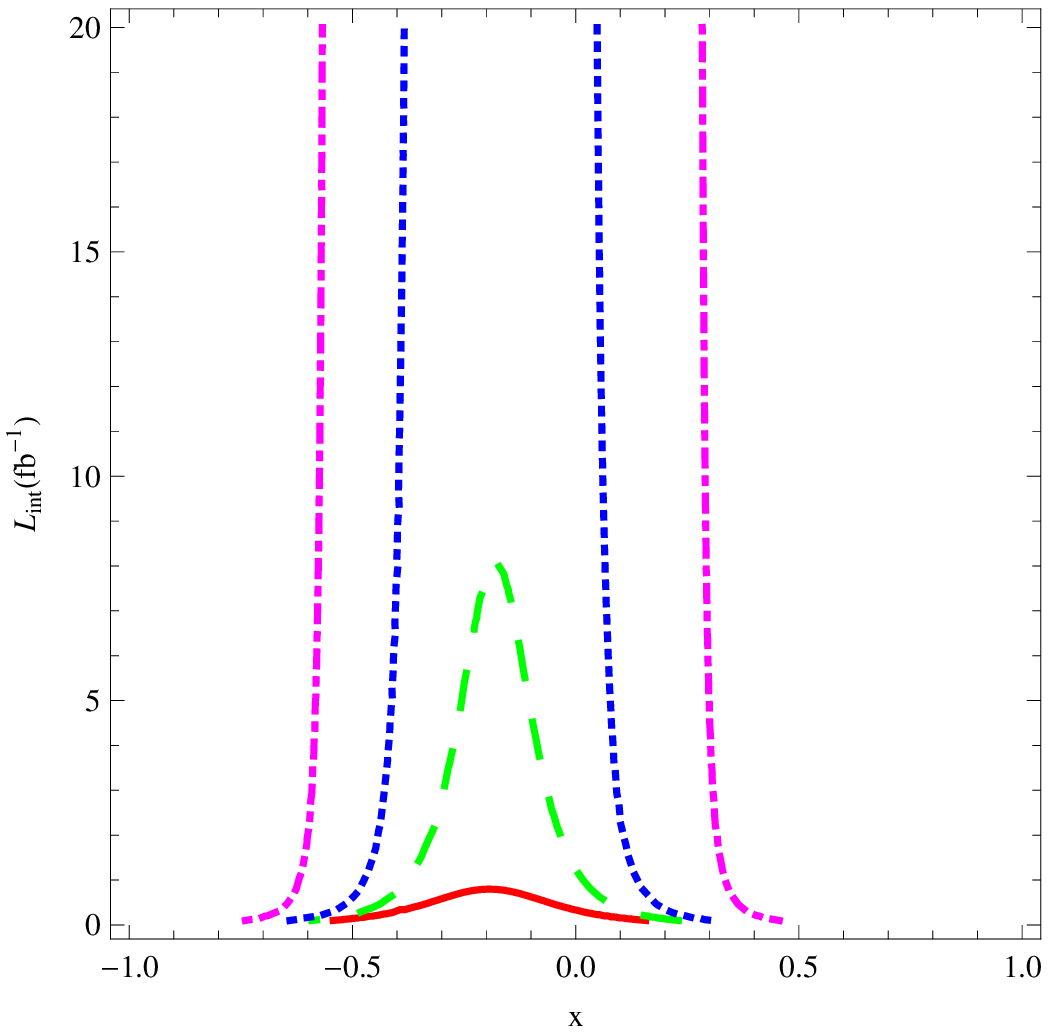}

\caption{The integrated luminosity needed to exclude ($95\%$ C.L.) the ranges
of the parameter $x$ at $\sqrt{s}=1$ TeV depending on the heavy
lepton masses: thick (red) line, dashed (green) line, dotted (blue)
line and dot-dashed (magenta) line corresponds to the mass values
$M_{l'}=350$, $375$, $400$ and $450$ GeV, respectively. \label{fig:fig7}}
\end{figure}

In order to present the potential of the linear colliders to search
for heavy lepton signal, we use a brief statistical significance analysis
with $S/\sqrt{B}$. We find accessible ranges of the mixing parameter
and the mass of heavy leptons depending on the integrated luminosity
at different center of mass energies. We present the exclusion plot
($95\%$ C.L.) for the heavy lepton mass $M_{l'}$ and mixing parameter
$x$ at $\sqrt{s}=1$ TeV and $L_{int}=20$, $100$, $1000$ pb$^{-1}$
and $10$ fb$^{-1}$ in Fig. \ref{fig:fig6}. Fig. \ref{fig:fig7}
presents the integrated luminosity needed to exclude ($95\%$ C.L.)
the ranges of the parameter $x$ at $\sqrt{s}=1$ TeV depending on
the heavy lepton masses of $350$, $375$, $400$ and $450$ GeV.

\begin{figure}
\includegraphics[scale=0.9]{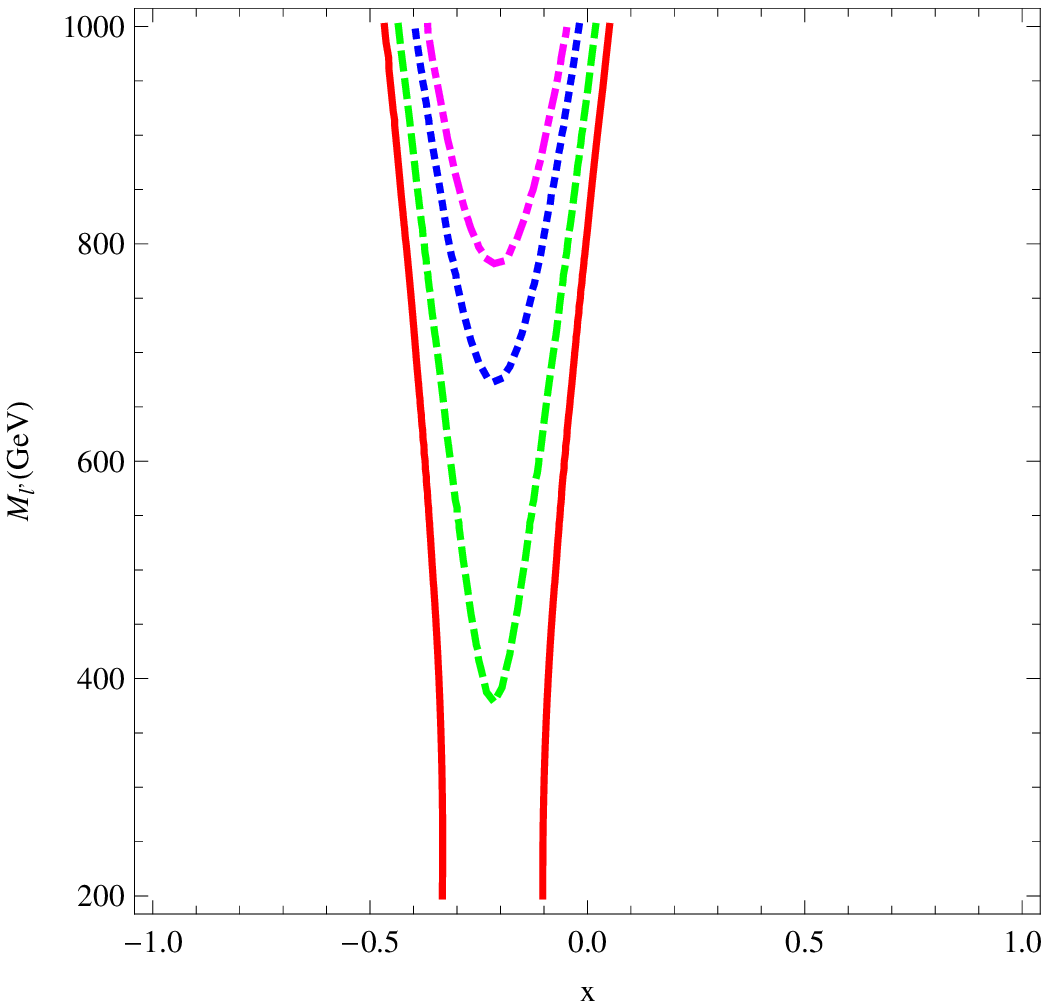}

\caption{Exclusion plot ($95\%$ C.L.) for the heavy lepton mass $M_{l'}$
and mixing parameter $x$ at $\sqrt{s}=3$ TeV and different integrated
luminosities: thick (red) line, dashed (green) line, dotted (blue)
line and dot-dashed (magenta) line corresponds to $L_{int}=5$, $10$,
$30$ and $100$ fb$^{-1}$, respectively. \label{fig:fig8}}
\end{figure}

\begin{figure}
\includegraphics[scale=0.9]{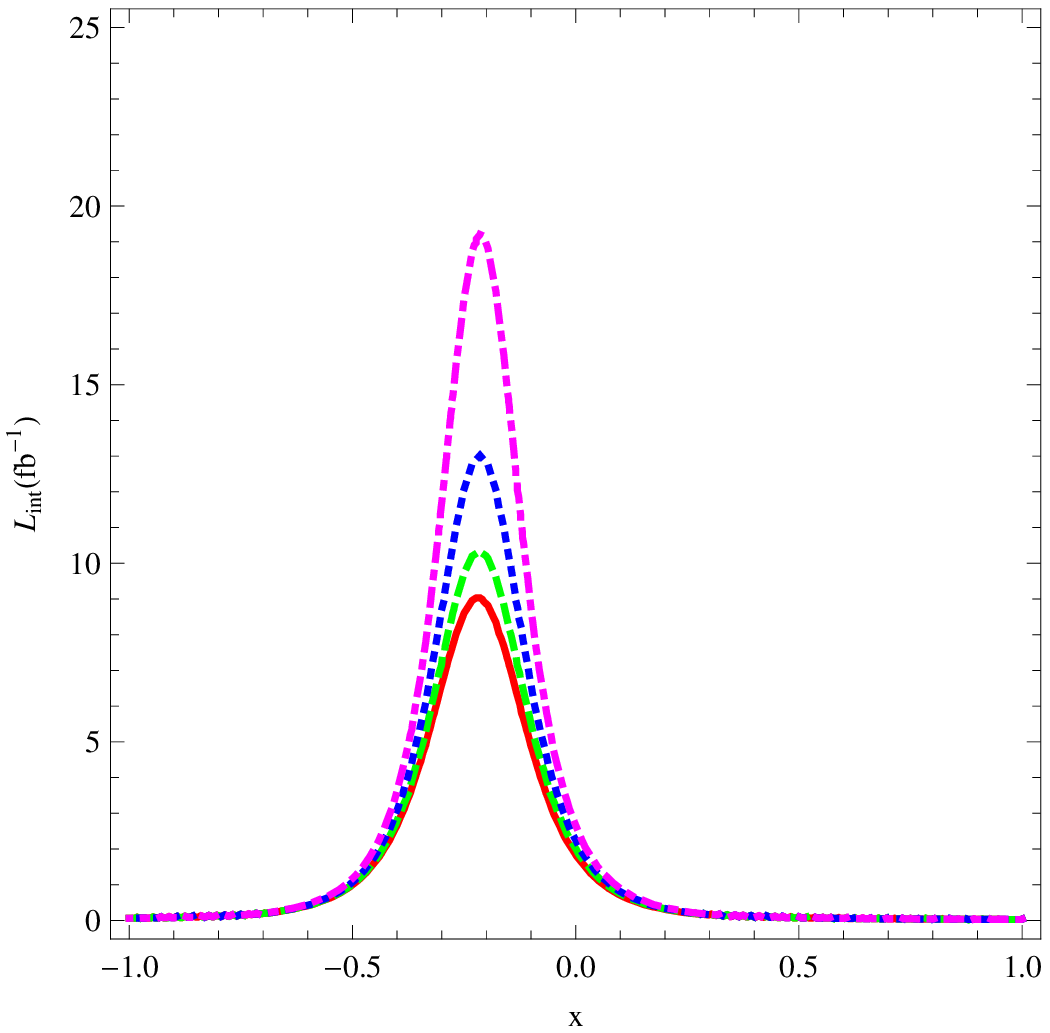}

\caption{The integrated luminosity needed to exclude ($95\%$ C.L.) the ranges
of the parameter $x$ at $\sqrt{s}=3$ TeV depending on the heavy
lepton masses: thick (red) line, dashed (green) line, dotted (blue)
line and dot-dashed (magenta) line corresponds to $M_{l'}=200$, $400$,
$500$ and $600$ GeV, respectively. \label{fig:fig9}}
\end{figure}

In Fig. \ref{fig:fig8}, we present the exclusion plot ($95\%$ C.L.)
for the heavy lepton mass $M_{l'}$ and mixing parameter $x$ at $\sqrt{s}=3$
TeV and $L_{int}=5$, $10$, $30$ and $100$ fb$^{-1}$. Fig. \ref{fig:fig9}
presents the integrated luminosity needed to exclude ($95\%$ C.L.)
the ranges of the parameter $x$ at $\sqrt{s}=3$ TeV depending on
the heavy lepton masses of $200$, $400$, $500$ and $600$ GeV.

It can be seen from the contour plots Fig. \ref{fig:fig6} and Fig.
\ref{fig:fig8}, the accessible range of the parameters are defined
outside of the contour lines depending on the luminosity of the collider.
For example, at $\sqrt{s}=1$ TeV and $L_{int}=100$ fb$^{-1}$ the
search can be performed beyond the range of mixing parameter $-0.4<x<0.1$
for given $M_{l'}=400$ GeV, while at $\sqrt{s}=3$ TeV and $L_{int}=10$
fb$^{-1}$ the inaccessible range becomes only $-0.3<x<-0.1$ for
given $M_{l'}=600$ GeV. We find the ranges of parameters which can
be accessed beyond $-0.4<x<0.1$ and $-0.25<x<-0.2$ for $M_{l'}=400$
GeV at the center of mass energies $\sqrt{s}=1$ TeV and $3$ TeV
at $L_{int}=10$ fb$^{-1}$, respectively. However, the whole region
of interest of the parameter $x$ can be accessed for $M_{l'}<800$
GeV at $\sqrt{s}=3$ TeV and $L_{int}=100$ fb$^{-1}$.

\section{Conclusion}

The charged $l'^{\pm}$ lepton will have a clear signature at linear
colliders. It is shown that the accessible ranges of the parameters
of new heavy leptons can be searched through the process $e^{+}e^{-}\to l'^{+}l'^{-}$
with their subsequent decays $l'^{\pm}\to\overset{(-)}{\nu}'+W^{\pm}$.
In this work, the lepton+dijet+missing transverse energy ($l^{\pm}+2j+$MET)
final states for the signal and background have been taken into account
to find the luminosity required to search for the new heavy lepton
mass and mixing parameter. If the LHC find clues about the new physics
models, the linear collider at TeV scale can enhance the accessible
range of parameter space from these models.
\begin{acknowledgments}
The work of O.C. and S.K. is supported in part by the Ministry of
Development (also formerly called State Planning Organization - DPT)
project under Grant No. DPT2006K-120470.\end{acknowledgments}

\end{document}